\documentstyle[12pt,psfig,graphicx]{article}                        

\begin{document}

\begin{center}
{\large {\bf MEASUREMENTS OF NUCLEAR LEVEL DENSITIES AND $\gamma$-RAY STRENGTH FUNCTIONS AND THEIR INTERPRETATIONS}}\\
~ \\

\underline{M.~Guttormsen}, M.~Hjorth-Jensen, J.~Rekstad and S.~Siem\\
Department of Physics, University of Oslo,
Box 1048 Blindern, N-0316 Oslo, Norway\\
~ \\
A.~Schiller\\
Lawrence Livermore National Laboratory, Livermore, CA-94551, USA\\ 
~ \\
A.~Voinov\\
Frank Laboratory of Neutron Physics, Joint Institute of Nuclear Research,
141980 Dubna, Moscow reg., Russia\\
\end{center}

\begin{abstract}
A method to extract primary $\gamma$-ray spectra from particle-$\gamma$ coincidences at excitation energies up to the neutron binding energy is described. From these spectra, the level density and $\gamma$-ray strength function can be determined. From the level density, several thermodynamical quantities are obtained within the microcanonical and canonical ensemble. Also models for the $\gamma$-ray strength function are discussed.

\end{abstract}

\section{Introduction}

The Oslo Cyclotron Group has established a method to deduce experimental level densities and $\gamma$-ray strength functions \cite{schi0}. These data contain essential information on nuclear structure and thermal and electromagnetic properties. In the last couple of years several fruitful applications of the method have been reported [2-10].

The most efficient way to create entropy in atomic nuclei is to break $J=0$ nucleon Cooper pairs in the core. A beautiful manifestation of pairbreaking is the backbending phenomena in rapidly rotating nuclei. In this work, the pairbreaking process is studied as function of intrinsic excitation energy (or temperature). Of great interest is the quenching of the pair correlations, a topic which is directly connected to the nuclear level density. 

Important applications of nuclear level densities and $\gamma$-ray strength functions are the determination of nuclear reaction cross sections from Hauser-Feshbach type of calculations. These cross sections are used as input parameters in large network calculations of stellar evolution, and in the simulation of accelerator-driven transmutation of nuclear waste.

\section{Experimental method and techniques}

The experiments were carried out using the ($^3$He,$\alpha \gamma$) and ($^3$He,$^3$He'$\gamma$) reactions at the Oslo Cyclotron Laboratory with a 45 MeV  $^3$He beam. Self-supporting metallic targets of rare earth nuclei were isotopically enriched to $\sim 95$\% and had thicknesses of $\sim 2$ mg/cm$^2$.

\begin{figure}[t]\centering
\mbox{\psfig{figure=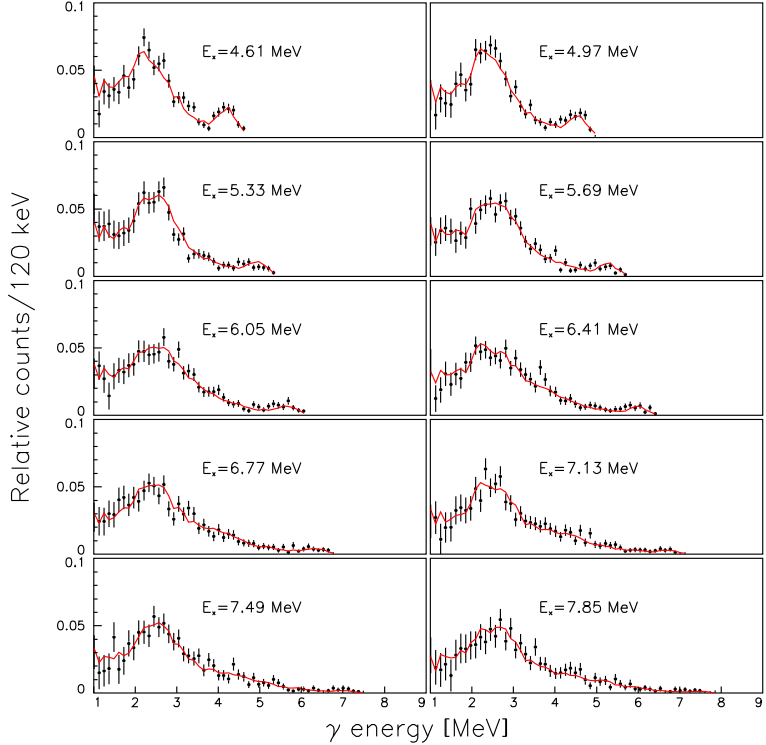,height=11.5cm}}
\caption{Comparison between observed primary $\gamma$ spectra for the $^{162}$Dy($^3$He,$^3$He'${\gamma})^{162}$Dy reaction (data points) and fits (lines) according to Eq.~(\ref{eq:ab}).}
\label{fig:fgboth}
\end{figure}

The charged particles and $\gamma$ rays were recorded with the detector array CACTUS, which contains eight particle telescopes and 27 NaI $\gamma$-ray detectors. Each telescope is placed at an angle of 45$^{\circ}$ relative to the beam axis, and comprises one Si front and one Si(Li) back detector with thickness $140$ and $3000$ {\normalfont $\mu$}m, respectively. The NaI $\gamma$-detector array, having a resolution of $\sim 6$ \% at $E_{\gamma} = 1$ MeV and a total efficiency of $\sim15$ \%, surrounds the target and particle detectors. In addition, two Ge detectors were used to monitor the spin distribution and the selectivity of the reactions.

In order to determine the true $\gamma$-energy distribution, the $\gamma$ spectra are corrected for the response of the NaI detectors with the unfolding procedure of Ref.~\cite{gutt4}. In addition random coincidences are subtracted from the $\gamma$ spectra. The set of unfolded $\gamma$ spectra are organized in a $(E,E_{\gamma})$ matrix, where the initial excitation energies $E$ are determined by means of reaction kinematics utilizing the energy of the ejectile. This matrix comprises the $\gamma$-energy distribution of the total $\gamma$ cascade. 

The primary $\gamma$ matrix can now be found according to the subtraction technique of Ref.~\cite{gutt0}, see data points of Fig.~\ref{fig:fgboth}. The procedure is based on the assumption that the decay properties of the particular reaction-selected distribution of states within each energy bin are independent on whether the respective ensembles of states are directly populated through the nuclear reaction or by $\gamma$-decay from higher lying states.

\begin{figure}[tbh]\centering
\mbox{\psfig{figure=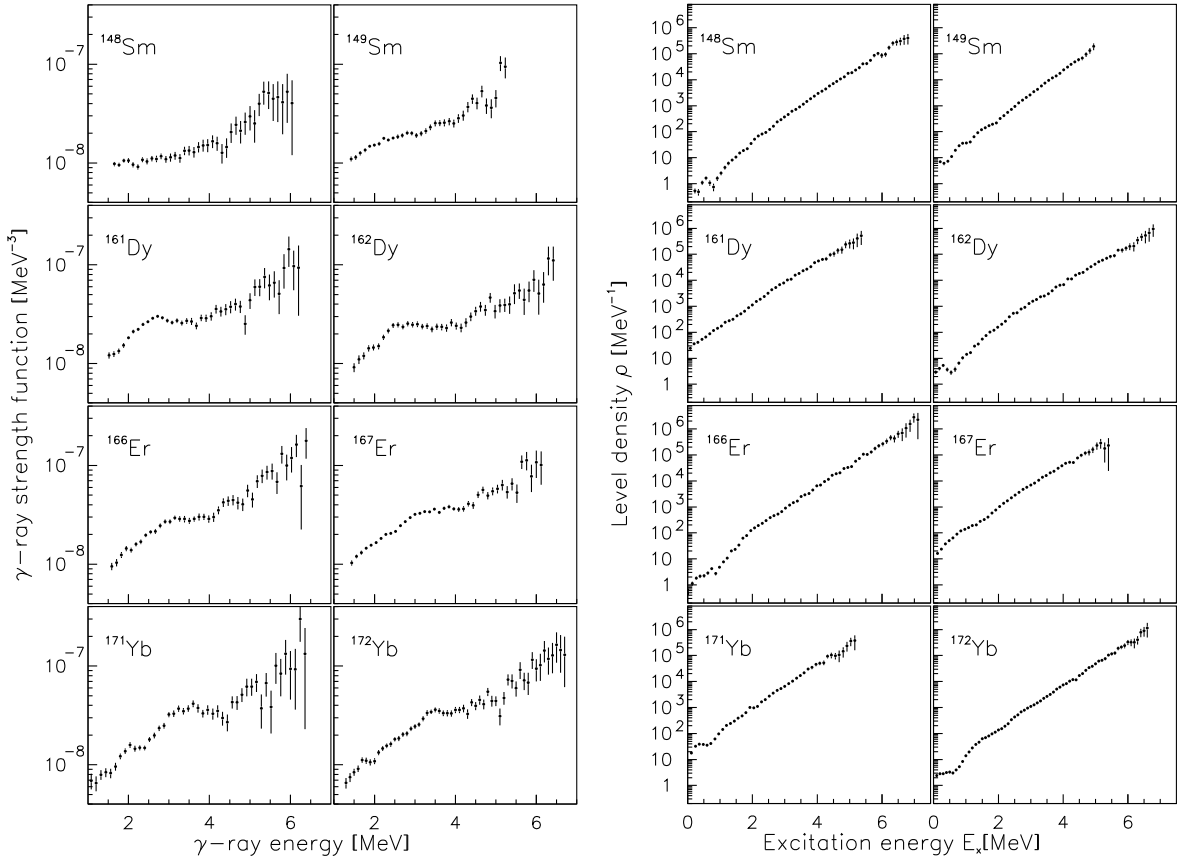,height=12.3cm}}
\caption{Observed $\gamma$-ray strength functions $f_{\rm E1+M1}$ and level densities $\rho$. The data may be downloaded from http://www.fys.uio.no/kjerne/english/cyclo/compilation.html.}
\label{fig:strfrho8}
\end{figure}

The idea is now to find two functions, the level density $\rho(E)$ and the $\gamma$-energy dependent function $F(E_{\gamma})$, that according to the Brink-Axel hypothesis~\cite{brink,axel}, should describe the primary $\gamma$-ray spectrum at excitation energy $E$:
\begin{equation}
P(E, E_{\gamma}) \propto \rho(E-E_{\gamma})F(E_{\gamma}).
\label{eq:ab}
\end{equation}
Here, $P(E, E_{\gamma})$ is fitted to the observed primary $\gamma$-ray matrix \cite{schi0}. 
In Fig.~\ref{fig:fgboth} the best fit to $P$ is shown for the $^{162}$Dy nucleus. All primary $\gamma$-ray spectra are seen to be very well described by the same $\rho$ and $F$ functions. However, from $\rho$ and $F$ one can construct other functions which give identical fits to the data by \cite{schi0}
\begin{eqnarray}
\tilde{\rho}(E-E_{\gamma}) &=& A \exp[\alpha(E-E_{\gamma})] \rho(E-E_{\gamma}) , 
\label{eq:array1}
\\
\tilde{F}(E_{\gamma}) &=& B \exp(\alpha E_{\gamma}) F(E_{\gamma}).
\label{eq:array2}
\end{eqnarray} 
Here, the parameters $A$ and $\alpha$ can be determined by fitting the level density to the number of known discrete levels at low excitation energy and to the level density estimated from neutron-resonance spacing data at high excitation energy. Furthermore, for dipole radiation, $F$ is proportional to $f{E_{\gamma}}^3$, where $f$ is the $\gamma$-ray strength function. Thus, according to Ref.~\cite{voin1}, the parameter $B$ can be determined from the known total $\gamma$ widths of neutron resonances. Finally, the normalized level densities and $\gamma$-ray strength functions are shown in Fig.~\ref{fig:strfrho8}.

\section{Thermodynamic properties}

High nuclear level density is due to the many ways the potential and kinetic energies of individual nucleons can add up to a certain excitation energy. Each valence nucleon, i.e.~nucleons not coupled in Cooper pairs, carries a single-particle entropy of $1.5-2.0$ in units of the Boltzmann constant $k_B$. Thus, a broken Cooper pair contributes significantly to the total entropy and heat capacity. Observed structures in the caloric curve and heat capacity may therefore indicate the breaking of Cooper pairs. 

\begin{figure}\centering
\mbox{\psfig{figure=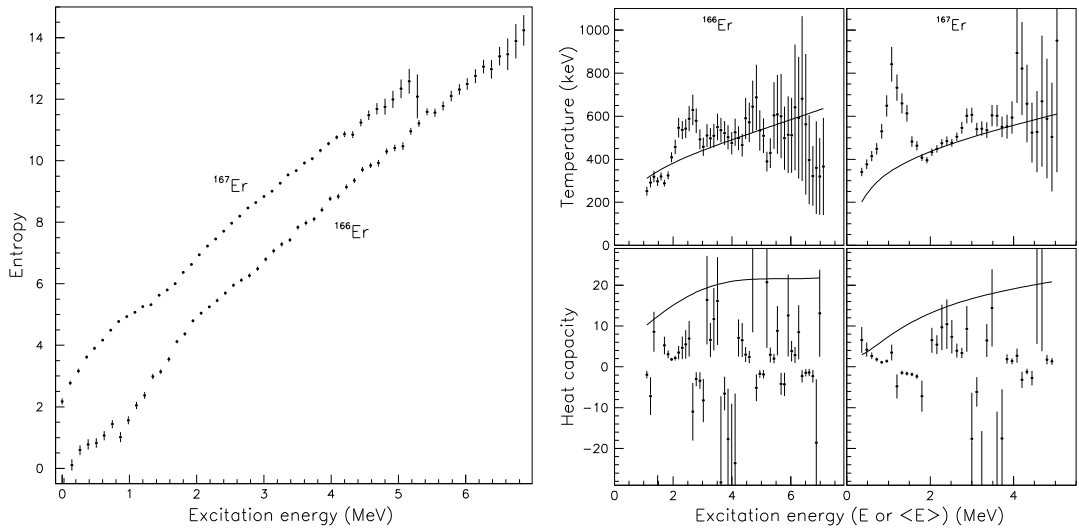,height=8cm}}
\caption{Left panel: The entropy as a function of excitation energy in $^{166}$Er and $^{167}$Er. Right panel: The temperature and heat capacity from the micro-canonical (data points) and canonical ensemble (lines).
}
\label{fig:termo}
\end{figure}

The microcanonical partition function is simply the multiplicity of nuclear states, which experimentally corresponds to the level density of accessible states. Thus, the experimental level density $\rho(E)$ is our starting point for the extraction of thermodynamic properties of nuclei.

The entropy in the micro-canonical ensemble is determined by
\begin{equation}
S(E)=\ln (\rho/\rho_0),
\end{equation}
where the Boltzmann constant is set to unity ($k_B\equiv 1$), and $\rho_0$ is adjusted to obtain $S\sim0$ in the ground-state band of even-even nuclei. The nuclear temperature and heat capacity is defined by
\begin{equation}
T(E)=\left(\frac{\partial S}{\partial E}\right)^{-1} {\rm and}
\label{eq:t}
\end{equation}
\begin{equation}
C_V(E) =\left(\frac{\partial T}{\partial E}\right)^{-1}.
\end{equation}
Figure~\ref{fig:termo} shows the entropy, temperature and heat capacity for $^{166}$Er and $^{167}$Er deduced in the microcanonical ensemble. The small bumps in the entropy curves are seen to be enhanced through the differentiations performed in Eq.~(\ref{eq:t}). The spectacular feature of negative branches in the heat capacity is a direct consequence of negative slopes in the caloric curve. Until today, their interpretation is a controversial topic. 

The canonical ensemble theory is based on the partition function 
\begin{equation}
Z(T)=\sum_{E=0}^{\infty}\omega (E)e^{-E/T},
\label{eq:z}
\end{equation}
which is a Laplace transform of the multiplicity of states $\omega (E)=\Delta E\rho (E)$, where $\Delta E$ is the energy bin.
\begin{figure}\centering
\mbox{\psfig{figure=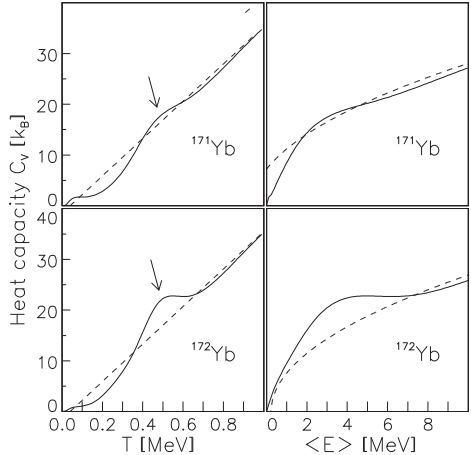,height=6cm}}
\caption{Canonical heat capacity as function of $T$ and $\langle E(T)\rangle$.}
\label{fig:cv}
\end{figure}
The thermal average of the excitation energy in the canonical ensemble is
\begin{equation}
\langle E(T)\rangle =Z^{-1}\sum_{E=0}^{\infty}E \omega (E)e^{-E/T}.
\label{eq:e}
\end{equation}
By the Laplace transform in Eq.~(\ref{eq:z}) much of the information contained in the micro-canonical level density is smeared out. Thus, fine structure in the thermodynamic observables in the microcanonical ensemble will not be visible in the canonical ensemble. The lines in Fig.~\ref{fig:termo} display the smooth dependence of the canonical temperature on $\langle E\rangle$. The corresponding heat capacity is the derivative by
\begin{equation}
C_V(T)=\frac{\partial \langle E\rangle}{\partial T},
\end{equation}
and is shown in Fig.~\ref{fig:cv} for $^{171,172}$Yb as a function of $T$ and $\langle E\rangle$. The heat capacities show a pronounced peak as function of temperature at $T_c=0.5$ MeV, which we interpret as the critical temperature for the pair breaking transition.

\section{Gamma-ray strength functions}

\begin{figure}\centering
\mbox{\psfig{figure=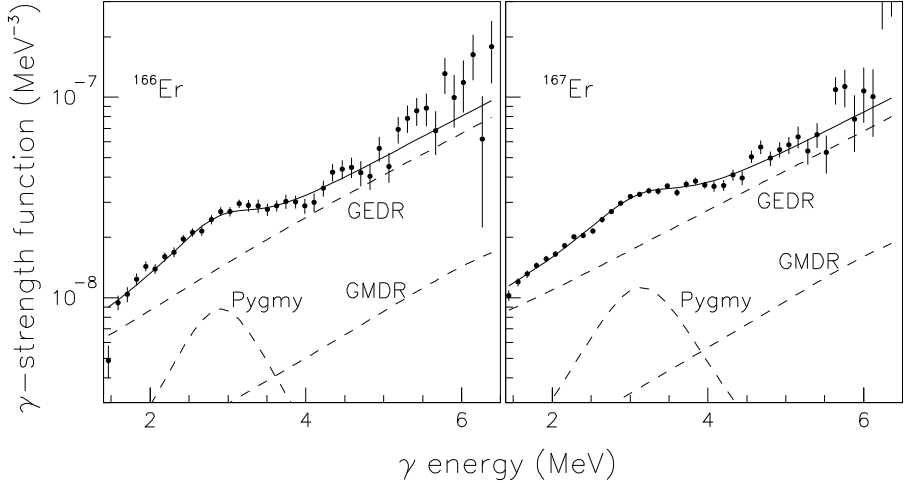,height=8cm}}
\caption{The experimental E1 + M1 $\gamma$-ray strength function (data points) of $^{166}$Er (left) and $^{167}$Er (right).
The solid line is the fit to the data by the theoretical model.
The dashed lines are the respective contributions of the GEDR, the GMDR, and the pygmy resonance to the total theoretical strength function.}
\label{fig:elin}
\end{figure}

\begin{figure}[htb]\centering
\mbox{\psfig{figure=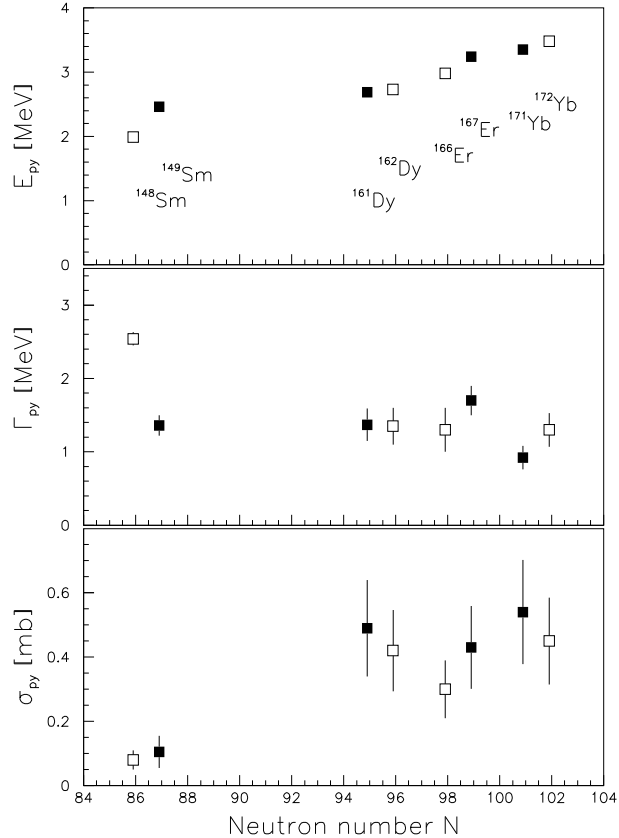,height=14cm}}
\caption{Systematics of the pygmy resonance parameters for odd (filled squares) and even (open squares) nuclei as a function of neutron number $N$. The resonance energy $E_{\mathrm{py}}$, the width $\Gamma_{\mathrm{py}}$ and the cross sections $\sigma_{\mathrm{py}}$ are shown in the upper, middle and lower panels, respectively.}
\label{fig:pygsys}
\end{figure}

In Fig.~\ref{fig:elin} the experimental $\gamma$-strength function is fitted by a theoretical strength function taking into account both the giant electric dipole resonance and the spin-flip resonance. In addition, a weaker resonance at lower energies is needed in order to fit the experimental data. Because of the much lower strength of this resonance compared to the giant electric dipole resonance (GEDR), it is denoted pygmy resonance. 
The E1 radiation is described by~\cite{kad}
\begin{equation} f_{\mathrm{E1}}(E_\gamma)=\frac{1}{3\pi^2\hbar^2c^2} \frac{0.7\sigma_{\mathrm{E1}}\Gamma_{\mathrm{E1}}^2(E_\gamma^2+4\pi^2T^2)} {E_{\mathrm{E1}}(E_\gamma^2-E_{\mathrm{E1}}^2)^2}, 
\end{equation}
where we apply $T$ as a constant fit parameter.
The M1 radiation is described by the Lorentzian
\begin{equation}
f_{\mathrm{M1}}(E_\gamma)=\frac{1}{3\pi^2\hbar^2c^2} \frac{\sigma_{\mathrm{M1}}E_\gamma\Gamma_{\mathrm{M1}}^2} {(E_\gamma^2-E_{\mathrm{M1}}^2)^2+E_\gamma^2\Gamma_{\mathrm{M1}}^2}, 
\label{eq:M1}
\end{equation}
where $\sigma_{\rm M1}$, $\Gamma_{\rm M1}$ and $E_{\rm M1}$ are giant magnetic dipole resonance (GMDR) parameters \cite{voin1}.
Furthermore, the pygmy resonance is described with a Lorentzian function $f_{\rm py}$ (similar as Eq.~(\ref{eq:M1})), where the pygmy-resonance strength $\sigma_{\rm py}$, width $\Gamma_{\rm py}$ and centroid $E_{\rm py}$ have been fitted in order to adjust the total theoretical strength function to the experimental data. The pygmy resonance parameters obtained for several rare earth nuclei are shown in Fig.~\ref{fig:pygsys}. The systematics indicate that the strengths $\sigma_{\mathrm{py}}$ of these soft dipole resonances are quenched when approaching the $N=82$ shell gap.

The pygmy resonance has been explained by the enhancement of the E1 $\gamma$-strength function \cite{iga}. Still, M1 character cannot be excluded. At an excitation energy around $3$ MeV, there is a concentration of orbital M1 strength in the weakly collective scissors mode~\cite{rich}. This mode was first observed in electron-scattering experiments~\cite{riht}, and is confirmed by the $(\gamma,\gamma ')$ reaction~\cite{pietralla}. Clarification of the electromagnetic character of the soft dipole resonance observed in our work awaits new experimental results.

\section{Conclusion}

Thermodynamic observables have been deduced from the level density and display signatures of phase-like transitions within the microcanonical and the canonical ensemble, interpreted as the transition from a strongly pair-correlated phase to an uncorrelated phase. In the microcanonical ensemble one may observe details of the successive breaking of nucleon pairs, information which is hidden in the canonical approach. The canonical ensemble on the other hand, reveals the average properties of the pairing transition. Using the canonical $C_V(T)$ quantity as thermometer, a local maximum is found at $T_c \sim 0.5$ MeV, indicating the breaking of Cooper pairs and quenching of pair correlations.

The experimental $\gamma$-strength function is fitted by a theoretical strength function, assuming that the $\gamma$ decay in the continuum is governed by dipole transitions. The contribution of electric and magnetic dipole radiation to the $\gamma$-strength function is recognized, and a bump is observed in the $\gamma$-strength function at $E_{\gamma}\sim 3$ MeV. A measurement of the electromagnetic character of this soft dipole resonance will be crucial in order to understand the physics behind this exciting phenomenon.

\section*{Acknowledgments}
Financial support from the Norwegian Research Council (NFR) is gratefully acknowledged. Part of this work was performed under the auspices of the U.S. Department of Energy by the University of California, Lawrence Livermore National Laboratory under Contract No. W-7405-ENG-48.

\end{document}